\RequirePackage{ifpdf}
\ifpdf % We are running pdfTeX in pdf mode
\documentclass[pdftex]{sigma}
\else
\documentclass{sigma}
\fi

\begin{document}
\allowdisplaybreaks

\renewcommand{\PaperNumber}{016}

\FirstPageHeading

\ShortArticleName{Extended Soliton Solutions in an Ef\/fective
Action for $SU(2)$ Yang--Mills Theory}

\ArticleName{Extended Soliton Solutions in an Ef\/fective Action\\
for $\boldsymbol{SU(2)}$ Yang--Mills Theory}

\Author{Nobuyuki SAWADO, Noriko SHIIKI and Shingo TANAKA}

\AuthorNameForHeading{N.~Sawado, N.~Shiiki and S.~Tanaka}
\Address{Department of Physics, Faculty of Science and Technology,
Tokyo University of Science,\\ Noda, Chiba 278-8510, Japan}

\Email{\href{mailto:sawado@ph.noda.tus.ac.jp}{sawado@ph.noda.tus.ac.jp},
 \href{mailto:norikoshiiki@mail.goo.ne.jp}{norikoshiiki@mail.goo.ne.jp}}

\ArticleDates{Received October 25, 2005, in f\/inal form January 25,
2006; Published online January 31, 2006}

\Abstract{The Skyrme--Faddeev--Niemi (SFN) model which is an
$O(3)$ $\sigma$ model in three dimensional space up to
fourth-order in the f\/irst derivative is regarded as a low-energy
ef\/fective theory of $SU(2)$ Yang--Mills theory. One can show from
the Wilsonian renorma\-li\-zation group argument that the ef\/fective
action of Yang--Mills theory recovers the SFN in the infrared
region. However, the theory contains an additional fourth-order
term which destabilizes the soliton solution. We apply the
perturbative treatment to the second derivative term in order to
exclude (or reduce) the ill behavior of the original action and
show that the SFN model with the second derivative term possesses
soliton solutions.}


\Classification{35Q51; 35G30; 70S15}

\section{\label{sec:level1}Introduction}

The Skyrme--Faddeev--Niemi (SFN) model which is an $O(3)$ $\sigma$
model in three dimensional space up to fourth-order in the f\/irst
derivative has topological soliton solutions with torus or
knot-like structure. The model was initiated in 70's and the
interest to it has been growing considerably. The numerical
simulations were performed
in~\cite{faddeev97,gladikowski97,sutcliffe98,hietarinta99,hietarinta00},
the integrability was shown in~\cite{aratyn99}, and the
application to the condensed matter physics~\cite{babaev02} and
the Weinberg--Salam model~\cite{fayzullaev} were also considered.
The recent research especially focuses on the consistency between
the SFN and fundamental theories such as
QCD~\cite{faddeev99,langmann99, shabanov99,cho02}. In those
references, it is claimed that the SFN action should be induced
from the $SU(2)$ Yang--Mills (YM) action at low energies. One can
also show from the Wilsonian renormalization group argument that
the ef\/fective action of Yang--Mills theory recovers the SFN in the
infrared region~\cite{gies01}. However, the derivative expansion
for slowly varying f\/ields $\boldsymbol{n}$ up to quartic order
produces an additional fourth-order term in the SFN model,
resulting in instability of the soliton solution.

Similar situations can be seen also in various topological soliton
models. In the Skyrme model, the chirally invariant Lagrangian
with quarks produces fourth order terms after the derivative
expansion and they destabilize the soliton
solution~\cite{dhar85,aitchison85}. To recover the stability of
the skyrmion, the author of~\cite{marleau01} introduced a large
number of higher order terms in the f\/irst derivative whose
coef\/f\/icients were determined from those of the Skyrme model by
using the recursion relations. Alternatively, in~\cite{gies01}
Gies pointed out the possibility that the second derivative order
term can work as a stabilizer for the soliton.

In this paper, we examine the Gies's supposition by numerical
analysis. In Section~\ref{sec:level2}, we give an introduction to
the Skyrme--Faddeev--Niemi model with its topological property. In
Section~\ref{sec:level3}, we show how to derive the SFN model
action from the $SU(2)$ Yang--Mills theory. In
Section~\ref{sec:level4}, soliton solutions of this truncated YM
action are studied. In order to f\/ind stable soliton solutions, we
introduce a second derivative term which can be derived in a
perturbative manner. The naive extremization scheme, however,
produces the fourth order dif\/ferential equation and the model has
no stable soliton solution. In Section~\ref{sec:level5}, the
higher derivative theory and Ostrogradski's formulation are
reviewed. We show the absence of bound state in the second
derivative theory using an example in quantum mechanics and
introduce the perturbative treatment for the second derivative
theory. In Section~\ref{sec:level6}, we present our numerical
results. Section~\ref{sec:level7} contains concluding remarks.

\section[Skyrme-Faddeev-Niemi model]{\label{sec:level2}Skyrme--Faddeev--Niemi model}

The Faddeev--Niemi conjecture for the low-energy model of $SU(2)$
Yang--Mills theory is expressed by the ef\/fective action:
\begin{gather}
S_{\rm SFN}=\Lambda^2 \int d^4x\left[\frac{1}{2}(\partial_\mu
\boldsymbol{n})^2 +\frac{g_1}{8}(\boldsymbol{n}\cdot \partial_\mu
\boldsymbol{n}\times \partial_\nu \boldsymbol{n})^2 \right] ,
\label{fsn_ac}
\end{gather}
where $\boldsymbol{n}(\boldsymbol{x})$ is a three component vector
f\/ield normalized as $\boldsymbol{n}\cdot\boldsymbol{n}=1$. The
mass scale parameter $\Lambda$ can be scaled out by the
replacement $\Lambda x \to x$ and $\Lambda^2 g_1 \to g_1$, and for
the static energy $E_{\rm stt}/\Lambda\to E_{\rm stt}$. Stable
soliton solutions exist when $g_1 > 0$.

The static f\/ield $\boldsymbol{n}(\boldsymbol{x})$ maps
$\boldsymbol{n}:R^3\mapsto S^2$ and the conf\/igurations are
classif\/ied by the topological maps characterized by a topological
invariant $H$ called Hopf charge
\begin{gather}
H=\frac{1}{32\pi^2}\int A \wedge F,\qquad F=dA, \label{hopf}
\end{gather}
where $F$ is the f\/ield strength and can be written as
$F=(\boldsymbol{n}\cdot d\boldsymbol{n}\wedge d\boldsymbol{n})$.

The static energy $E_{\rm stt}$ from the action (\ref{fsn_ac}) has
a topological lower bound~\cite{vakulenko},
\begin{gather}
E_{\rm stt}\ge K H^{3/4}, \label{lowerbound}
\end{gather}
where $K=4\sqrt{2}~3^{3/8}\pi^2\sqrt{g_1}$. Note that Ward
improved this topological bound by using the Hopf
map~\cite{ward98}. It seems, however, to be an upper bound of the
model rather than lower bound.

Performing numerical simulation, one can f\/ind that the static
conf\/igurations for $H=1,2$ have axial symmetry~\cite{sutcliffe98}.
Thus ``the toroidal ansatz'' which was studied
in~\cite{gladikowski97} is suitable to be imposed on these
conf\/igurations. The ansatz is given by
\begin{gather}
n_1=\sqrt{1-w^2(\eta,\beta)}\cos(N\alpha+v(\eta,\beta)), \nonumber \\
n_2=\sqrt{1-w^2(\eta,\beta)}\sin(N\alpha+v(\eta,\beta)), \nonumber \\
n_3=w(\eta,\beta),\label{toroidal}
\end{gather}
where $(\eta,\beta,\alpha)$ are toroidal coordinates which are
related to the $R^3$ as follows:
\begin{gather*}
x=\frac{a\sinh\eta\cos\alpha}{\tau},\qquad
y=\frac{a\sinh\eta\sin\alpha}{\tau},\qquad
z=\frac{a\sin\beta}{\tau}
\end{gather*}
with $\tau=\cosh\eta-\cos\beta$.

The function $w(\eta,\beta)$ is subject to the boundary conditions
$w(0,\beta)=1$, $w(\infty,\beta)=-1$ and is periodic in $\beta$.
$v(\eta,\beta)$ is set to be
$v(\eta,\beta)=M\beta+v_0(\eta,\beta)$ and $v_0(,\beta)$ is
considered as a constant map. Equation (\ref{hopf}) then gives
$H=NM$.

In this paper we adopt a simpler ansatz than (\ref{toroidal}),
which is def\/ined by
\begin{gather}
n_1=\sqrt{1-w^2(\eta)}\cos(N\alpha+M\beta),\nonumber \\
n_2=\sqrt{1-w^2(\eta)}\sin(N\alpha+M\beta),\nonumber  \\
n_3=w(\eta),  \label{afz}
\end{gather}
where $w(\eta)$ satisf\/ies the boundary conditions $w(0)=1$,
$w(\infty)=-1$. We numerically study soliton solutions for both
ansatz (\ref{toroidal}) and (\ref{afz}). By comparing those
results, we f\/ind that this simple ansatz produces at most 10\,\%
errors and does not much ef\/fect to the property of the soliton
solution.

By using (\ref{afz}), the static energy is written in terms of the
function $w(\eta)$ as
\begin{gather*}
E_{\rm stt}=2\pi^2a \int d\eta \left[ \frac{(w')^2}{1-w^2}+(1-w^2)
U_{M,N}(\eta)+\frac{g_1}{4a^2}\sinh\eta\cosh\eta (w')^2
U_{M,N}(\eta)\right], \\
w'\equiv \frac{dw}{d\eta},\qquad U_{M,N}(\eta)\equiv
\left(M^2+\frac{N^2}{\sinh^2\eta}\right).
\end{gather*}
The Euler--Lagrange equation of motion is then derived as
\begin{gather}
\frac{w''}{1-w^2}+\frac{ww'^2}{(1-w^2)^2}+U_{M,N}(\eta)w
+\frac{g_1}{2a^2}\bigl(-2N^2\coth^2\eta w'
U_{M,N}(\eta)w' \nonumber \\
\qquad {}+\big(\cosh^2\eta+\sinh^2\eta\big)+\sinh\eta\cosh\eta
U_{M,N}(\eta)w''\bigr)=0. \label{fsn_eq}
\end{gather}
The variation with respect to $a$ produces the equation for
variable $a$. Soliton solutions are obtained by solving the
equations for $a$ as well as for $w$.

\section[Effective action in the Yang-Mills theory
with CFNS decomposition]{\label{sec:level3}Ef\/fective action in the Yang--Mills theory\\
with CFNS decomposition}

In this section, we brief\/ly review how to derive the SFN ef\/fective
action from the action of SU(2) Yang--Mills theory in the infrared
limit~\cite{shabanov99,gies01}. For the gauge f\/ields
$\boldsymbol{A}_\mu$, the Cho--Faddeev--Niemi--Shabanov
decomposition is
applied~\cite{faddeev99,langmann99,shabanov99,cho02}
\begin{gather}
\boldsymbol{A}_\mu=\boldsymbol{n}C_\mu+(\partial_\mu\boldsymbol{n})\times\boldsymbol{n}+\boldsymbol{W}_\mu.
\label{cfns}
\end{gather}
The f\/irst two terms are the ``electric'' and ``magnetic'' Abelian
connection, and $\boldsymbol{W}_\mu$ are chosen to be orthogonal
to $\boldsymbol{n}$, i.e.\
$\boldsymbol{W}_\mu\cdot\boldsymbol{n}=0$. Obviously, the degrees
of freedom on the left- and right-hand side of
equation~(\ref{cfns}) do not match. While the LHS describes
$3_{\rm color}\times 4_{\rm Lorentz}=12$, the RHS is comprised of
$(C_\mu:)4_{\rm Lorentz}+(\boldsymbol{n}:)2_{\rm
color}+(\boldsymbol{W}_\mu:)3_{\rm color} \times4_{\rm
Lorentz}-4_{\boldsymbol{n}\cdot\boldsymbol{W}_\mu=0}=14$ degrees
freedom. Shabanov introduced in his paper \cite{shabanov99} the
following constraint
\begin{gather*}
\boldsymbol{\chi}(\boldsymbol{n},C_\mu,\boldsymbol{W}_\mu)=0,\qquad
{\rm with}\quad \boldsymbol{\chi}\cdot\boldsymbol{n}=0.
\end{gather*}
The generating functional of YM theory can be written by using
equation~(\ref{cfns}) as
\begin{gather*}
{\cal Z}=\int {\cal D}\boldsymbol{n}{\cal D}C{\cal
D}\boldsymbol{W}\delta(\boldsymbol{\chi}) \Delta_{\rm
FP}\Delta_{\rm S}e^{-S_{\rm YM}-S_{\rm gf}}.
%\label{vf0}
\end{gather*}
$\Delta_{\rm FP}$ and $S_{\rm gf}$ are the Faddeev--Popov
determinant and the gauge f\/ixing action respectively, and Shabanov
introduced another determinant $\Delta_{\rm S}$ corresponding to
the condition $\boldsymbol{\chi}=0$. YM and the gauge f\/ixing
action is given by
\begin{gather*}
S_{\rm YM}+S_{\rm  gf}=\int
d^4x\left[\frac{1}{4g^2}\boldsymbol{F}_{\mu\nu}\cdot\boldsymbol{F}_{\mu\nu}
+\frac{1}{2\alpha_{\rm g} g^2}(\partial_\mu
\boldsymbol{A}_\mu)^2\right] .
\end{gather*}
Inserting equation~(\ref{cfns}) into the action, one obtains the
form:
\begin{gather*}
{\cal Z}=\int {\cal D}\boldsymbol{n}e^{-{\cal S}_{\rm
ef\/f}(\boldsymbol{n})} =\int {\cal D}\boldsymbol{n} e^{-{\cal
S}_{\rm cl}(\boldsymbol{n})}\int {\cal D}\tilde{C}{\cal
D}\boldsymbol{W}_\mu
\Delta_{\rm FP}\Delta_{\rm S}\delta({\boldsymbol \chi}) \nonumber \\
\phantom{{\cal Z}=\int {\cal D}\boldsymbol{n}e^{-{\cal S}_{\rm
ef\/f}(\boldsymbol{n})}=}{} \times e^{-(1/2g^2) \int(\tilde{C}_\mu
M^C_{\mu\nu}\tilde{C}_\nu+\boldsymbol{W}_\mu
\bar{M}^{\boldsymbol{W}}_{\mu\nu}\boldsymbol{W}_\nu +2C_\nu
K^C_\nu+2\boldsymbol{W}_\mu\cdot\boldsymbol{K}^{\boldsymbol{W}}_\mu)}
\end{gather*}
with
\begin{gather*}
M^C_{\mu\nu}=-\partial^2\delta_{\mu\nu}+\partial_\mu\boldsymbol{n}\cdot\partial_\nu\boldsymbol{n}, \nonumber \\
M^{\boldsymbol{W}}_{\mu\nu}=-\partial^2\delta_{\mu\nu}-\partial_\mu\boldsymbol{n}\otimes\partial_\nu\boldsymbol{n}
+\partial_\nu\boldsymbol{n}\otimes\partial_\mu\boldsymbol{n}, \nonumber \\
\boldsymbol{Q}^C_{\mu\nu}=\partial_\mu\boldsymbol{n}\partial_\nu+\partial_\nu\boldsymbol{n}\partial_\mu
+\partial_\mu\partial_\nu\boldsymbol{n},\nonumber \\
K^C_{\mu\nu}=\partial_\nu(\boldsymbol{n}\cdot\partial_\nu\boldsymbol{n}\times\partial_\mu\boldsymbol{n})
+\partial_\mu\boldsymbol{n}\cdot\partial^2\boldsymbol{n}\times\boldsymbol{n}, \nonumber \\
\boldsymbol{K}^{\boldsymbol
W}_{\mu\nu}=\partial_\mu(\boldsymbol{n}\times\partial^2\boldsymbol{n}),
\qquad {\rm (in~gauge~\alpha_g=1})
\end{gather*}
and
\begin{gather*}
\bar{M}^{\boldsymbol{W}}_{\mu\nu}:=M^{\boldsymbol{W}}_{\mu\nu}+\tilde{\boldsymbol{Q}}^C_{\mu
s}
{M^C}^{-1}_{s\lambda}\boldsymbol{Q}^C_{\lambda \nu}, \nonumber \\
\tilde{C}_\mu=C_\mu+\boldsymbol{W}_s\cdot
\boldsymbol{Q}_{s\lambda}{M^C}^{-1}_{\lambda\mu}.
\end{gather*}
Here, $\tilde{\boldsymbol{Q}}^C_{\mu s}$ has same form of
$\boldsymbol{Q}^C_{\mu s}$ but dif\/ferentiates to the right objects
of ${M^C}^{-1}_{s\lambda}\boldsymbol{Q}^C_{\lambda \nu}$. The
classical action of $\boldsymbol{n}$ including the gauge f\/ixing
term is given by
\begin{gather*}
{\cal S}_{\rm cl}=\int
d^4x\left[\frac{1}{4g^2}(\partial_\mu\boldsymbol{n}\times
\partial_\nu\boldsymbol{n})^2 +\frac{1}{2 \alpha_{\rm g}
g^2}\big(\partial^2\boldsymbol{n}\times\boldsymbol{n}\big)^2\right].
\end{gather*}
The $\delta$ functional is expressed by its Fourier transform
\begin{gather*}
\delta(\boldsymbol{\chi})=\int {\cal D}\boldsymbol{\phi}e^{-i\int
(\boldsymbol{\phi}\cdot \partial\boldsymbol{W}_\mu
+\boldsymbol{\phi}\cdot C_\mu\boldsymbol{n}\times
\boldsymbol{W}_\mu
+(\boldsymbol{\phi}\cdot\boldsymbol{n})(\partial_\mu\boldsymbol{n}\cdot\boldsymbol{W}_\mu))}.
\end{gather*}
Integrating over $C$, $\boldsymbol{W}$, $\boldsymbol{\phi}$, we
f\/inally obtain
\begin{gather}
e^{-S_{\rm ef\/f}}=e^{-S_{\rm cl}}\Delta_{\rm FP}\Delta_{\rm S}
\big(\det M^C\big)^{-1/2} \big(\det
\bar{M}^{\boldsymbol{W}}\big)^{-1/2} \big(\det
-\tilde{Q}^{\boldsymbol{\phi}}_\mu
\big(\bar{M}^{\boldsymbol{W}}\big)^{-1}_{\mu\nu}
Q^{\boldsymbol{\phi}}_\nu\big)^{-1/2}, \label{determ}
\end{gather}
where
$Q^{\boldsymbol{\phi}}_\nu=i(-\partial_\mu+\partial_\mu\boldsymbol{n}\otimes\boldsymbol{n})$
and $\tilde{Q}^{\boldsymbol{\phi}}_\mu$ dif\/ferentiates to the
right. We perform the derivative expansion for the four
determinants in equation~(\ref{determ}) under the following
assumptions
\begin{itemize}
\itemsep=0pt \item [(i)]the theory is valid for the momenta $p$
with $k<p<\Lambda$ ($k,\Lambda$ are infrared and ultraviolet
cut-of\/f), \item [(ii)]$|\partial\boldsymbol{n}| \ll k$, \item
[(iii)]the higher derivative terms, such as
$\partial^2\boldsymbol{n}$ are omitted.
\end{itemize}
The ef\/fective action is then given by
\begin{gather}
S_{\rm ef\/f}=\int d^4x\left[\frac{1}{2}(\partial_\mu
\boldsymbol{n})^2 +\frac{g_1}{8}(\partial_\mu \boldsymbol{n}\times
\partial_\nu \boldsymbol{n})^2 +\frac{g_2}{8}(\partial_\mu
\boldsymbol{n})^4
 \right].
\label{fsn2}
\end{gather}
For $g_1>0$ and $g_2=0$, the action is identical to the FSN
ef\/fective action~(\ref{fsn_ac}).

In order to get the stable soliton solutions, $g_2$ must be
positive~\cite{gladikowski97}. However, $g_2$ is found to be
negative according to the above analysis (see~\cite{gies01}).
Therefore we consider higher-derivative terms and investigate if
the model with the higher-derivatives possess soliton solutions.

\section{\label{sec:level4}Search for the stable soliton solutions (1)}
The static energy is derived from equation~(\ref{fsn2}) as
\begin{gather*}
E_{\rm stt}=  \int d^3x\left[\frac{1}{2}(\partial_i
\boldsymbol{n})^2 +\frac{g_1}{8}(\partial_i \boldsymbol{n}\times
\partial_j \boldsymbol{n})^2
+\frac{g_2}{8}(\partial_i \boldsymbol{n})^4 \right] \nonumber \\
\phantom{E_{\rm
stt}}{}:=E_2(\boldsymbol{n})+E_4^{(1)}(\boldsymbol{n})+E_4^{(2)}(\boldsymbol{n}).
\end{gather*}
A spatial scaling behavior of the static energy, so called
Derrick's scaling argument, can be applied to examine the
stability of the soliton~\cite{sutcliffe05}. Considering the map
$\boldsymbol{x}\mapsto \boldsymbol{x}'=\mu\boldsymbol{x}$
$(\mu>0)$, with $\boldsymbol{n}^{(\mu)}\equiv
\boldsymbol{n}(\mu\boldsymbol{x})$, the static energy scales as
\begin{gather}
e(\mu) = E_{\rm stt}\big(\boldsymbol{n}^{(\mu)}\big) \nonumber \\
\phantom{e(\mu)}{}=E_2\big(\boldsymbol{n}^{(\mu)}\big)+
E_4^{(1)}\big(\boldsymbol{n}^{(\mu)}\big)+E_4^{(2)}\big(\boldsymbol{n}^{(\mu)}\big) \nonumber \\
\phantom{e(\mu)}{}=\frac{1}{\mu}E_2(\boldsymbol{n})+\mu\big(E_4^{(1)}(\boldsymbol{n})+E_4^{(2)}(\boldsymbol{n})\big).
\label{derrick}
\end{gather}
Derrick's theorem states that if the function $e(\mu)$ has no
stationary point, the theory has no static solutions of the f\/ield
equation with f\/inite density, other than the vacuum. Conversely,
if~$e(\mu)$ has stationary point, the  possibility of having
f\/inite energy soliton solutions is not excluded.
equation~(\ref{derrick}) is stationary at
$\mu=\sqrt{E_2/\big(E_4^{(1)}+E_4^{(2)}\big)}$. Then, the
following inequality
\begin{gather*}
g_1(\partial_i\boldsymbol{n}\times\partial_j\boldsymbol{n})^2
+g_2(\partial_i\boldsymbol{n})^2(\partial_j\boldsymbol{n})^2 \nonumber \\
\qquad{}=g_1(\partial_i\boldsymbol{n})^2(\partial_j\boldsymbol{n})^2
-g_1(\partial_i\boldsymbol{n}\cdot\partial_j\boldsymbol{n})^2
+g_2(\partial_i\boldsymbol{n})^2(\partial_j\boldsymbol{n})^2 \nonumber \\
\qquad {} \geqq
g_2(\partial_i\boldsymbol{n}\cdot\partial_j\boldsymbol{n})^2
\qquad \big(\because
(\partial_i\boldsymbol{n})^2(\partial_j\boldsymbol{n})^2 \geqq
(\partial_i\boldsymbol{n}\cdot\partial_j\boldsymbol{n})^2\big)
\end{gather*}
ensures the possibility of existence of the stable soliton
solutions for $g_2\geqq 0$. As mentioned in the
Section~\ref{sec:level3}, $g_2$ should be negative at least within
our derivative expansion analysis of YM theory.

A promising idea to tackle the problem was suggested by
Gies~\cite{gies01}. He considered the following type of ef\/fective
action, accompanying second derivative term
\begin{gather}
S_{\rm ef\/f}=\int d^4x\left[\frac{1}{2}(\partial_\mu
\boldsymbol{n})^2 +\frac{g_1}{8}(\partial_\mu \boldsymbol{n}\times
\partial_\nu \boldsymbol{n})^2 -\frac{g_2}{8}(\partial_\mu
\boldsymbol{n})^4 +\frac{g_2}{8}\big(\partial^2
\boldsymbol{n}\cdot\partial^2 \boldsymbol{n}\big) \right].
\label{fsn_ac2}
\end{gather}
Here we choose positive value of $g_2$ and assign the explicit
negative sign to the third term. In principle, it is possible to
estimate the second derivative term by the derivative expansion
without neglecting throughout the calculation.

The static energy of equation~(\ref{fsn_ac2}) with the ansatz
(\ref{afz}) is written as
\begin{gather*}
E_{\rm stt}=2\pi^2a \int d\eta \left[ \frac{(w')^2}{1-w^2}+(1-w^2)
U_{M,N}(\eta)
+\frac{g_1}{4a^2}\sinh\eta\cosh\eta (w')^2 U_{M,N}(\eta)\right.\nonumber \\
\phantom{E_{\rm
stt}=}{}+\frac{g_2}{4a^2}\left[-\sinh\eta\cosh\eta\left[\frac{(w')^2}{1-w^2}
+(1-w^2)U_{M,N}(\eta)\right]^2\right. \nonumber \\
\phantom{E_{\rm
stt}=}{}+\bigl(\coth\eta+\sinh^2\eta-\sinh\eta\cosh\eta\bigr)
\frac{(w')^2}{1-w^2} \nonumber \\
\phantom{E_{\rm
stt}=}{}+\big(\sinh\eta\cosh\eta-\sinh^2\eta\big)\big(1-w^2\big)M^2
+2\left\{ \frac{w(w')^3}{(1-w^2)^2} +\frac{w' w''}{1-w^2}
+w w'U_{M,N}(\eta) \right\} \nonumber \\
\phantom{E_{\rm stt}=}{}+\left.\left.\sinh\eta\cosh\eta\left\{
\frac{1}{1-w^2}\left[\frac{(w')^2}{1-w^2} +w w''
+(1-w^2)U_{M,N}(\eta)\right]^2 +(w'')^2\right\}\right]\right],
\end{gather*}
where $w''\equiv \frac{d^2w}{d\eta^2}$. The Euler--Lagrange
equation of motion is derived by
\begin{gather*}
-\frac{d^2}{d\eta^2}\left(\frac{\partial E_{\rm stt}}{\partial
w''}\right) +\frac{d}{d\eta}\left(\frac{\partial E_{\rm
stt}}{\partial w'}\right) -\frac{\partial E_{\rm stt}}{\partial
w}=0 ,
\end{gather*}
which is too complicated to write down explicitly and hence we
adopt the following notation
\begin{gather}
f_0(w,w',w'')+g_1f_1(w,w',w'')+g_2f_2\big(w,w',w'',w^{(3)},w^{(4)}\big)=0.
\label{fsn_eq2}
\end{gather}
Here $w^{(3)}$, $w^{(4)}$ represent the third and the fourth
derivative with respect to~$\eta$. The f\/irst two terms of
equation~(\ref{fsn_eq2}) are identical to those in
equation~(\ref{fsn_eq}). Unfortunately, we could not f\/ind out
stable soliton solutions from equation~(\ref{fsn_eq2}) for any
(even in the quite small) value of $g_2$.

From the relation
\begin{gather*}
\int d^4x\big[\big(\partial^2 \boldsymbol{n}\cdot\partial^2
\boldsymbol{n}\big)-(\partial_\mu \boldsymbol{n})^4\big] =\int
d^4x\big(\partial^2 \boldsymbol{n}\times\boldsymbol{n}\big)^2,
\end{gather*}
one easily f\/inds that the static energy obtained from the last two
terms in equation~(\ref{fsn_ac2})
\begin{gather}
\tilde{E}^{(2)}_4= \int d^3x\big(\partial^2
\boldsymbol{n}\times\boldsymbol{n}\big)^2 \label{energy2}
\end{gather}
gives the positive contribution. The total static energy is
stationary at
$\mu=\sqrt{E_2/\big(E_4^{(1)}+\tilde{E}_4^{(2)}\big)}$ and hence
the possibility of existence of soliton solutions is not excluded.
And also, the positivity of equation~(\ref{energy2}) does not
spoil the lower bound~(\ref{lowerbound}) of original SFN and the
possibility still remains.

The standard action (\ref{fsn_ac}) has a fourth dif\/ferential order
term, but only quadratic in time derivative. On the other hand,
the action (\ref{fsn_ac2}) involves a term with fourth order in
time derivative. Then, even if the static energy is constructed
only by the positive terms, it is in general not bounded from
below. As a result, the stability of the soliton is unclear. In
the next section, we report the basic feature of the higher
derivative action and introduce the method to avoid the problems
inherent by applying an example in quantum mechanics.

\section{\label{sec:level5}Higher derivative theory}

In this section, we make a small detour, i.e., we review the
problems in the higher derivative
theory~\cite{pais,smilga,eliezer89,jaen,simon} which essentially
falls into two categories. The f\/irst problem concerns the increase
in the number of degrees of freedom. For example, if the theory
contains second derivative terms, the equation of motion becomes
up to the order in the fourth derivative. Thus, four parameters
are required for the initial conditions. If one considers
higher-order terms, the situation gets worse. However, this is not
a serious problem for our study because our concern is existence
of static soliton solutions. The second problem is that the
actions of the theory are not bounded from below. This feature
makes the higher derivative theories unstable.

We brief\/ly review the Lagrangian and the Hamiltonian formalism
with higher derivative called the Ostrogradski method. We consider
the Lagrangian containing up to $n$th order derivatives
\begin{gather*}
S=\int dt {\cal L}\big(q,\dot{q},\ldots,q^{(n)}\big).
\end{gather*}
Taking the variation of the action $\delta S=0$ leads the
Euler--Lagrange equation of motion
\begin{gather*}
\sum_{i=0}^n (-1)^i\frac{d^i}{dt^i}\left(\frac{\partial {\cal
L}}{\partial q^{(i)}}\right)=0.
\end{gather*}
The Hamiltonian is obtained by introducing $n$ generalized momenta
\begin{gather*}
p_i=\sum_{j=i+1}^n
(-1)^{j-i-1}\frac{d^{j-i-1}}{dt^{j-i-1}}\left(\frac{\partial {\cal
L}}{\partial q^{(j)}}\right), \qquad i=1,\ldots,n,
\end{gather*}
or
\begin{gather}
p_n=\frac{\partial {\cal L}}{\partial q^{(n)}},\qquad
p_i=\frac{\partial {\cal L}}{\partial
q^{(i)}}-\frac{d}{dt}p_{i+1},\qquad i=1,\ldots,n-1,
\label{canonical momenta}
\end{gather}
and $n$ independent variables
\begin{gather*}
q_1\equiv q,\qquad q_i\equiv q^{(i-1)},\qquad i=2,\ldots,n.
\end{gather*}
The Lagrangian now depends on the $n$ coordinates $q_i$ and on the
f\/irst derivative $\dot{q}_n=q^{(n)}$. The Hamiltonian is def\/ined
as
\begin{gather*}
{\cal H}(q_i,p_i)=\sum^n_{i=1}p_i\dot{q}_i-{\cal
L}=\sum^{n-1}_{i=1} p_i q_{i+1}+p_n \dot{q}_n-{\cal L}.
\end{gather*}
The canonical equations of motion turn out to be
\begin{gather*}
\dot{q}_i=\frac{\partial {\cal H}}{\partial p_i},\qquad
\dot{p}_i=-\frac{\partial {\cal H}}{\partial q_i}.
\end{gather*}
Thus, we replace a theory of one coordinate $q$ system obeying
$2n$-th dif\/ferential equation with a set of 1-st order canonical
equations for $2n$ phase-space variables $[q_i,p_i]$.

We consider a simple example including second derivative
term~\cite{simon}, def\/ined as
\begin{gather*}
{\cal
L}=\frac{1}{2}\big(1+\varepsilon^2\omega^2\big)\dot{q}^2-\frac{1}{2}\omega^2q^2-\frac{1}{2}\varepsilon^2\ddot{q}^2,
\end{gather*}
where constant $\epsilon$ works as a coupling constant of second
derivative term. The equation of motion~is
\begin{gather}
\big(1+\varepsilon^2 \omega^2\big)\ddot{q}+\omega^2
q+\varepsilon^2 q^{(4)}=0. \label{eq_quanta}
\end{gather}
From equation~(\ref{canonical momenta}), one gets
\begin{gather*}
\pi_{\dot{q}}=\frac{\partial{\cal L}}{\partial
\ddot{q}}=-\varepsilon^2\ddot{q},\qquad \pi_q=\frac{\partial{\cal
L}}{\partial \dot{q}}-\frac{d}{dt}\left(\frac{\partial{\cal
L}}{\partial \ddot{q}}\right)
=\big(1+\varepsilon^2\omega^2\big)\dot{q}+\varepsilon^2\dddot{q}.
\end{gather*}
Thus the Hamiltonian becomes
\begin{gather*}
{\cal H}=\dot{q}\pi_q+\ddot{q}\pi_{\dot{q}}-{\cal L} \nonumber \\
\phantom{{\cal
H}}{}=\dot{q}\pi_q-\frac{1}{2\varepsilon^2}\pi_{\dot{q}}^2
-\frac{1}{2}\big(1+\varepsilon^2\omega^2\big)\dot{q}^2+\frac{1}{2}\omega^2q^2.
\end{gather*}
We introduce the new canonical variables
\begin{gather*}
q_+=\frac{1}{\omega\sqrt{1-\varepsilon^2\omega^2}}\big(\varepsilon^2\omega^2\dot{q}-\pi_q\big),\qquad
p_+=\frac{w}{\sqrt{1-\varepsilon^2\omega^2}}(q-\pi_{\dot{q}}), \nonumber \\
q_-=\frac{\varepsilon}{\sqrt{1-\varepsilon^2\omega^2}}(\dot{q}-\pi_q),\qquad
p_-=\frac{1}{\varepsilon\sqrt{1-\varepsilon^2\omega^2}}\big(\varepsilon^2\omega^2
q-\pi_{\dot{q}}\big),
\end{gather*}
and the Hamiltonian will have the following form by using these
variables
\begin{gather*}
{\cal H}\to \frac{1}{2}\big(p_+^2+\omega^2
q_+^2\big)-\frac{1}{2}\left(p_-^2+\frac{1}{\varepsilon^2}
q_-^2\right).
\end{gather*}
The corresponding energy spectra is then given by
\begin{gather}
E=\left(n+\frac{1}{2}\right)\omega-\left(m+\frac{1}{2}\right)\frac{1}{\varepsilon},\qquad
n,m=0,1,2,\ldots. \label{qenergy}
\end{gather}
One can see that in the limit $\epsilon\to 0$ the energy goes to
negative inf\/inity rather than approaching to the harmonic
oscillator energy eigenstates.

To obtain physically meaningful solutions, we employ the
perturbative analysis where the solution is expanded in terms of
the small coupling constant and the Euler--Lagrange equation of
motion is replaced with the corresponding perturbative equation.
The solutions of the equations of motion that are ill behaved in
the limit $\epsilon\to 0$ are excluded from the very
beginning~\cite{eliezer89,jaen,simon}.

We assume that the solution of equation~(\ref{eq_quanta}) can be
written as
\begin{gather}
q_{\rm pert}(t)=\sum^{\infty}_{n=0} \epsilon^n q(t). \label{q}
\end{gather}
Substituting equation~(\ref{q}) into equation~(\ref{eq_quanta})
and taking time derivatives of these equations, we obtain the
constraints for higher derivative terms
\begin{gather}
O\big(\epsilon^0\big)  \nonumber \\
\quad {\rm equation:}\quad \ddot{q}_0+\omega^2q_0=0, \nonumber\\
\quad {\rm constraints:}\quad \dddot{q}_0=-\omega^2\dot{q}_0, \qquad \ddddot{q}_0=\omega^4 q_0. \label{cnst0} \\
O\big(\epsilon^2\big)  \nonumber \\
\quad {\rm equation:}\quad \ddot{q}_2+\omega^2\ddot{q}_0+\omega^2 q_2+\ddddot{q}_0=0,  \nonumber \\
\phantom{\quad {\rm equation:}\quad}{}\Rightarrow \ddot{q}_2+\omega^2q_2=0,\qquad ({\rm using}~(\ref{cnst0})),\nonumber\\
\quad {\rm constraints:}\quad \dddot{q}_2=-\omega^2\dot{q}_2,\qquad  \ddddot{q}_2=\omega^4 q_2. \label{cnst2} \\
O\big(\epsilon^4\big)  \nonumber \\
\quad {\rm equation:}\quad \ddot{q}_4+\omega^2\ddot{q}_2+\omega^2 q_4+\ddddot{q}_2=0,  \nonumber \\
\phantom{\quad {\rm equation:}\quad}{}\Rightarrow \ddot{q}_4+\omega^2q_4=0,\qquad ({\rm using}~(\ref{cnst2})),\nonumber\\
\quad {\rm constraints:}\quad \dddot{q}_4=-\omega^2\dot{q}_4, \qquad \ddddot{q}_4=\omega^4 q_4. \nonumber%\label{cnst4}
\end{gather}
Combining these results, we f\/ind the perturbative equation of
motion up to $O\big(\epsilon^4\big)$
\begin{gather*}
\ddot{q}_{\rm pert}+\omega^2q_{\rm pert}=O\big(\epsilon^6\big),
\end{gather*}
which is the equation for harmonic oscillator.

\begin{figure}[t]
\begin{minipage}[t]{7.5cm}
\centerline{\includegraphics[width=7.5cm]{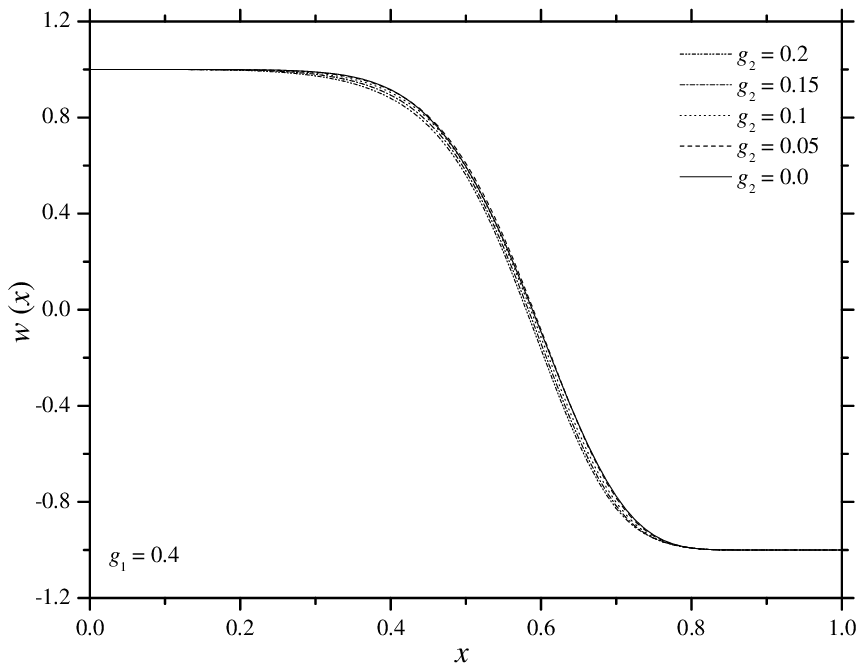}}
\vspace{-3mm} \caption{\label{fig:Fig1} The function $w(\eta)$ for
$g_1=0.4$, $g_2=0,\,0.05,\,0.1,\,0.15,\,0.2$ (the rescaling radial
coordinate $x=\eta/(1-\eta)$ is used).}
\end{minipage}
\qquad
\begin{minipage}[t]{7.5cm}
\centerline{\includegraphics[width=7.5cm]{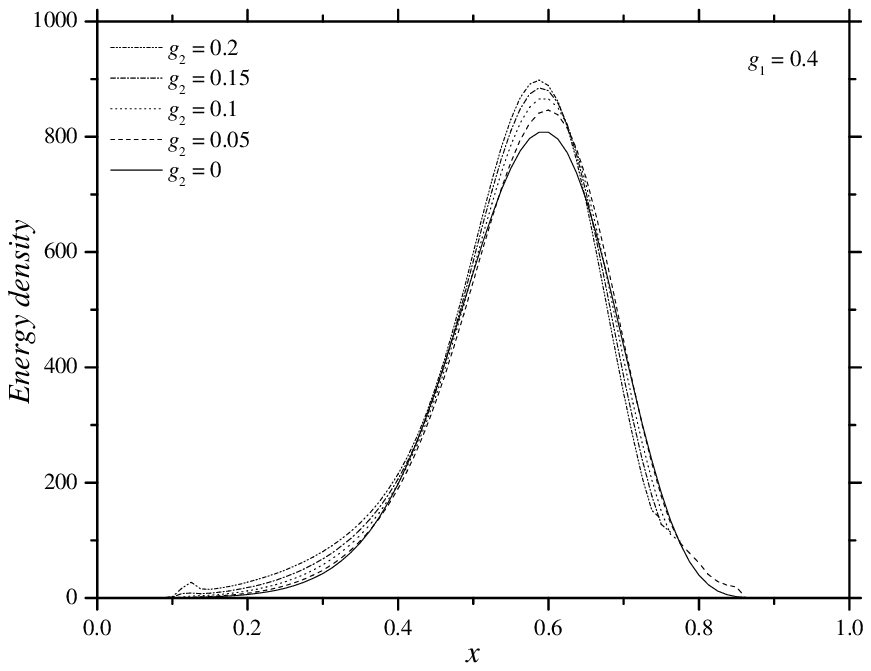}}
\vspace{-3mm} \caption{\label{fig:Fig2} The energy density for
$g_1=0.4$, $g_2=0,\,0.05,\,0.1,\,0.15,\,0.2$.}
\end{minipage}
\end{figure}

\begin{figure}[th]
\centerline{\includegraphics[width=7.5cm]{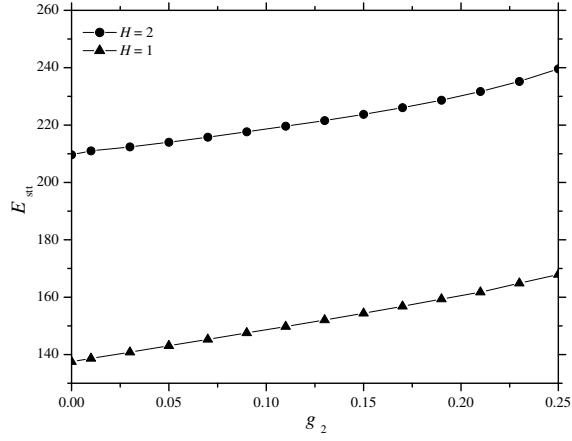}}
\vspace{-3mm} \caption{\label{fig:Fig3} The energy as a function
of $g_2$ ($g_1=0.4$).}
\end{figure}

This perturbative method can successfully exclude the ill behavior
of the second derivative theory. Of course our problem is
dif\/ferent from this quantum mechanics example. While the energy
spectra~(\ref{qenergy}) is quantum value, the energy of the
soliton is static and classical. Nevertheless, the perturbative
method may give some hint to tackle our problem. In the next
section, we will apply the method to the action with the second
derivative term~(\ref{fsn_ac2}) and explore the stable soliton
solutions.

\section{\label{sec:level6}Search for the stable soliton solutions (2) \\
(perturbative expansion method)}

As in the case of quantum mechanics, we assume that $g_2$ is
relatively small and can be considered as a perturbative coupling
constant. Thus, in analogy of (\ref{q}), the perturbative solution
is written by a power series in $g_2$
\begin{gather}
w(\eta)=\sum^\infty_{n=0}g_2^n w_n(\eta). \label{sol_expand}
\end{gather}
Substituting equation~(\ref{sol_expand}) into
equation~(\ref{fsn_eq2}), we obtain the classical f\/ield equation
in $O(g^0_2)$
\begin{gather}
f_0(w_0,w_0',w_0'')+g_1f_1(w_0,w_0',w_0'')=0. \label{classical}
\end{gather}
Taking derivatives for both sides in equation~(\ref{classical})
and solving for $w_0''$, we obtain the following form of the
constraint equations for higher derivatives
\begin{gather}
w_0^{(i)}=F^{(i)}(w_0,w_0'),\qquad i=2,3,4. \label{constraint}
\end{gather}
The equation in $O\big(g^1_2\big)$ can be written as
\begin{gather}
(f_0+g_2
f_1)_{O(g^1_2)}+f_2\big(w_0,w_0',w_0'',w_0^{(3)},w_0^{(4)}\big)=0.
\label{fsn_eq21}
\end{gather}
Substituting the constraint equations (\ref{constraint}) into
equation~(\ref{fsn_eq21}) and eliminating the higher derivative
terms, one can obtain the perturbative equation of motion
\begin{gather}
f_0(w,w',w'')+g_1 f_1(w,w',w'')+g_2
\tilde{f}_2(w,w')=O\big(g^2_2\big). \label{fsn_eq2_p}
\end{gather}
Now equation~(\ref{fsn_eq2_p}) has stable soliton solutions.

Our results of the estimated function $w(\eta)$ and the energy
density are displayed in Figs.~\ref{fig:Fig1} and~\ref{fig:Fig2}.
(In all f\/igures, we show the results for the case of Hopf charge
$H=2$; $N=2$, $M=1$.) The dependence of the total energy on~$g_2$
is shown in Fig.~\ref{fig:Fig3}. The change is moderate with
respect to~$g_2$.

\section{\label{sec:level7}Summary}

In this paper we have studied the Skyrme--Faddeev--Niemi model and
its extensions by introducing the reduction scheme of the $SU(2)$
Yang--Mills theory to the corresponding low-energy ef\/fective
model. The requirement of consistency between the low-energy
ef\/fective action of the YM and the SFN type model leads us to take
into account second derivative terms in the action. However, we
found that such an action including the second derivative terms
does not have stable soliton solutions. This is due to the absence
of the energy bound in higher derivative theory. This fact
inspired us to employ the perturbative analysis to our ef\/fective
action. Within the perturbative analysis, we were able to obtain
the stable soliton solutions.

Our analysis is based on perturbation and the coupling constant
$g_2$ is assumed to be small. However, Wilsonian renormalization
analysis of YM theory~\cite{gies01} suggests that the coupling
constants~$g_1$, $g_2$ (and the mass scale parameter $\Lambda$)
depend on the renormalization group time $t=\log k/\Lambda$
($k,\Lambda$ are infrared, ultraviolet cutof\/f parameter) and those
are almost comparable. To improve the analysis, we could perform
the next order of perturbation, but it is tedious and spoils the
simplicity of the SFN model.

It should be noted that our solutions do not dif\/fer much from the
solution of original SFN model, at least in the perturbative
regime. We suspect that an appropriate truncation (such as ``extra
fourth order term + second derivative term'') always supplies the
stable solutions that are close to the original SFN model. Thus we
conclude that the topological soliton model comprised of the
``kinetic term + a special fourth order term'' like SFN model is a
good approximation.

Since our results were obtained from numerical study in the
perturbative approach, it is uncertain whether the existence of
the soliton is kept for larger coupling constant $g_2$. To conf\/irm
that, we should proceed to investigate next order perturbation,
or, otherwise, f\/ind some analy\-ti\-cal evidence for that. We
point out that the perturbative treatment is only used for
excluding the ill behavior of the second derivative f\/ield theory.
We hope that applying this prescription does not alter the
essential feature of the solutions.

Finally, let us mention the application of the soliton solutions
to the glueball. Obviously this is one of the main interests to
study the model and, many authors have given discussions on this
subject~\cite{gies01,faddeev04,cho04}. On the other hand, the
possibility of the magnetic condensation of the QCD vacuum within
the Cho--Faddeev--Niemi--Shabanov decomposed Yang--Mills theory
have been studied by Kondo~\cite{kondo}. The author claims the
existence of nonzero of\/f diagonal gluon mass $M_X$, which is
induced in terms of the condensation of the magnetic potential
part of the decomposition $\boldsymbol{B}_\mu\sim
(\partial_\mu\boldsymbol{n})\times\boldsymbol{n}$, as
\begin{gather*}
M_X^2=\langle
\boldsymbol{B}_\mu\cdot\boldsymbol{B}_\mu\rangle=\langle
 (\partial_\mu\boldsymbol{n})^2\rangle.
 \end{gather*}
Throughout our calculation, we scaled out the mass scale parameter
$\Lambda$ in the action (\ref{fsn_ac}) but, in this sense, it
should ref\/lect the information of such gluon mass, or the
condensation property of the vacuum. After a careful examination
of the value of $\Lambda$, we will be able to make a prediction
for the glueball mass.

\subsection*{Acknowledgements}
We are grateful to Kei-Ichi Kondo for drawing our attention to
this subject and for many useful advises. We also thank
M.~Hirayama and J.~Yamashita for valuable discussions.

\LastPageEnding

\end{document}